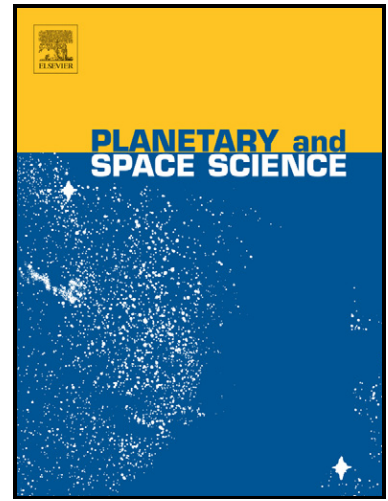

Measurement requirements for a near-earth asteroid impact mitigation demonstration mission

Stephen D. Wolters, Andrew J. Ball, Nigel Wells, Christopher Saunders, Neil McBride

Cite this article as: Stephen D. Wolters, Andrew J. Ball, Nigel Wells, Christopher Saunders and Neil McBride, Measurement requirements for a near-earth asteroid impact mitigation demonstration mission, *Planetary and Space Science*, doi:10.1016/j.pss.2011.06.015

# Measurement Requirements for a Near-Earth Asteroid Impact Mitigation Demonstration Mission

Stephen D. Wolters[a], Andrew J. Ball[b], Nigel Wells[c], Christopher Saunders[c], and Neil McBride[a]

Corresponding author: Stephen D. Wolters. Tel: +44 01908 659465. e-mail: s.d.wolters@open.ac.uk.

[a] Planetary and Space Sciences Research Institute, Centre for Earth, Planetary, Space and Astronomical Research, The Open University, Walton Hall, Milton Keynes MK7 6AA, UK
[b] Former Postdoctoral Research Fellow, PSSRI, The Open University. Present address: Nieuwe Rijn 28D, 2312 JD Leiden ZH, The Netherlands.
[c] QinetiQ, Cody Technology Park, Farnborough, Hampshire GU14 0LX, UK

## Abstract

A concept for an Impact Mitigation Preparation Mission, called *Don Quijote*, is to send two spacecraft to a Near-Earth Asteroid (NEA): an Orbiter and an Impactor. The Impactor collides with the asteroid while the Orbiter measures the resulting change in the asteroid's orbit, by means of a Radio Science Experiment (RSE) carried out before and after impact. Three parallel Phase A studies on *Don Quijote* were carried out for the European Space Agency: the research presented here reflects outcomes of the study by QinetiQ. We discuss the mission objectives with regards to the prioritisation of payload instruments, with emphasis on the interpretation of the impact. The Radio Science Experiment is described and it is examined how solar radiation pressure may increase the uncertainty in measuring the orbit of the target asteroid. It is determined that to measure the change in orbit accurately a thermal IR spectrometer is mandatory, to measure the Yarkovsky effect. The advantages of having a laser altimeter are discussed. The advantages of a dedicated wide-angle impact camera are discussed and the field-of-view is initially sized through a simple model of the impact.

## 1 Introduction

Deflection of a threatening Near-Earth Asteroid (NEA) by kinetic impact is prominent among possible mitigation methods examined in recent years. The principle is conceptually simple, i.e. that an impulse applied to an NEA well in advance of the predicted impact can modify its orbital parameters sufficient to avoid a collision (e.g. Ahrens & Harris 1992). In 2002 the *Don Quijote* mission concept was proposed, with the aim of demonstrating the capability to deliver an impactor to an NEA and measure the effect on its orbit, in the context of the NEO Space Mission Preparation activity of ESA's General Studies Programme (Milani et al. 2003). It was then refined further in the ESA Concurrent Design Facility (CDF) (Carnelli et al. 2006), prior to parallel Phase A studies carried out by industrial consortia in 2006/2007 [e.g. Rathke & Izzo (2006), which concentrates on modelling the orbital dynamics of the impact].

ESA specified two mission objectives for the DQ Phase A studies:

**Primary Objective:** Change the asteroid's semi-major axis by > 100 m and measure the change to 1% accuracy. Determine momentum transfer from the impact by measuring the asteroid's mass, size, bulk density and rotation state.

**Secondary Objective:** perform multi-spectral mapping of the asteroid.

Two primary target asteroids were specified for the studies: 2002 $AT_4$ and (10302) 1989 ML. Their key properties are given in Table 1 (note that some parameters are imposed 'engineering estimates' rather than confirmed values from observations). 2002 $AT_4$, has a smaller diameter (~320 m) and hence lower mass, but a larger semi-major axis and higher eccentricity ($a$ = 1.866 AU, $e$ = 0.447), making it more difficult to rendezvous with ($\Delta v$ ~6.6 km s$^{-1}$) but easier to achieve a large deflection. It also has an interesting primitive D-type composition. In contrast, (10302) 1989 ML has a larger diameter (680 m), smaller semi-major axis and less eccentric orbit ($a$ = 1.272 AU, $e$ = 0.136), making it easier to rendezvous with ($\Delta v$ ~4.5 km s$^{-1}$) but harder to deflect.

Three parallel Phase A studies on *Don Quijote* were carried out for the European Space Agency. This paper focuses on the definition and prioritisation of the payload and science operations for the QinetiQ-led study, highlighting aspects relevant more generally to any similar mission concepts. These include:

- The necessity of an 'Impact Interpretation Objective' in addition to the Primary and Secondary objectives listed above, and consequent measurement requirements for the payload;
- The need for thermal IR measurements of the target asteroid;
- The significance of the uncertainty in solar radiation pressure on the Orbiter spacecraft.

The Executive Summary of the Phase A study can be downloaded from http://esa-mm.esa.int/docs/NEO/QinetiqDQExecSum.pdf. The DQ mission profile is shown in Figure 1. Two separate launches deliver first an Orbiter and then an Impactor to the target asteroid. The Orbiter allows precise determination of the target's orbit both before and after the collision of the Impactor spacecraft with the target, by means of a Radio Science Experiment. For the baseline scenarios identified in the study, a semi-major axis change of 3000 m was predicted for 2002 AT$_4$, and 90 m for (10302) 1989 ML.

## 2 Mission Objectives, Measurement Requirements and Payload Instruments

Table 2 shows that each of the mission objectives gives rise to a set of top-level measurement requirements. While the Primary and Secondary Objectives would together result in a mission that demonstrates a measured deflection and yields valuable multispectral imagery of the target asteroid, this study concluded that the benefit of such a mission for future NEO mitigation can only be realised by adopting a *verification* approach. This demands knowledge of the conditions under which deflection was achieved, i.e. key features of the target material and impact event. Without such an approach, planning for a future impact mitigation mission would have to cope with considerable uncertainty in the degree of scaling that would be required with respect to the DQ design parameters. The main reason for this is that the momentum enhancement factor of the impact event depends on properties of the target as well as those of the impactor. The momentum gained by the asteroid is the direct momentum transfer of the impactor, plus the momentum arising from ejecta thrown back at speeds greater than escape speed. Different hydrocode approaches currently produce wildly different results for different material properties, extending to factors as high as 35 (Benz & Nyffeler 2004) For example, an analysis by Holsapple (2004) using a wave code with a three-phase equation of state found that for non-porous asteroids the momentum multiplication may be around a factor of 4, while for a porous asteroid it could be as low as 1.1. This study thus assumed a worst-case momentum enhancement factor of 1 (i.e. no enhancement).

Adopting this verification approach, we thus proposed an additional objective, intermediate in priority to those specified by ESA. The **Impact Interpretation Objective** is to calibrate the impact through measurement of relevant near-surface properties and observation of the impact, so that the momentum enhancement factor that results from the impact can be extrapolated more precisely for the general population of NEAs. As a consequence, the payload needs to include instruments that determine:

- the near-surface bulk density/porosity;
- the near-surface grain size;
- the impact ejecta mass distribution;
- the ejecta speed distribution.

Table 2 lists proposed instrument types that would address one or more of the top-level measurement requirements arising from all three mission objectives. Heritage and possible resource allocations are shown in Table 3.

A thorough physical characterisation of the target asteroid would enable the measurement of the momentum change to be placed in context with the general population of NEAs and the much greater database of ground-based observations. Analysis from Orbiter camera images would provide geophysical and geomorphological characterisation. This can be combined with mineralogical composition from NIR spectrometry and elemental abundances from X-ray spectroscopy that provide a compositional picture for comparison with meteorites. While the asteroid macroporosity will be estimated from its bulk density, the mineralogical composition obtained with NIR and X-ray spectrometry can be correlated with meteorite types to estimate microporosity (e.g. Britt & Consolmagno 2003). Thermal characterization using a thermal IR spectrometer will make it possible to measure precisely the thermal inertia at different points on the surface. This parameter is crucial for modelling the semi-major axis drift arising from the Yarkovsky effect. This drift must be distinguished from the impulsive change in the semi-major axis due to the impact.

The following sections address each of the payload experiments in turn, highlighting the key features.

## 3 Radio Science Experiment (RSE)

The RSE would (combined with data from the Orbiter Camera and a Laser Altimeter) determine the mass, centre of mass (as opposed to centre of figure) and low-order gravity field harmonics, both before and after the impact. It would thus help determine the change in semi-major axis of the asteroid's orbit as a result of the impact. The motion of the Orbiter is monitored using Doppler shifts in the frequencies of Ka/Ka-band (and also the X/X-band) transponder system and solving for the orbit and gravitational field of the asteroid, with knowledge of the relative position of the asteroid and Orbiter, and the asteroid's rotation state. It may also be possible, over time, to measure the secular change in orbital parameters due to the Yarkovsky effect.

### *3.1 Initial Drift-bys*
We assume the RSE has an initial phase with hyperbolic drift-bys for an initial estimate of the NEA mass (before orbit insertion). The accuracy of a fly-by mass determination can be estimated from (Patzold et al. 2001):

$$\frac{\sigma_{GM}}{GM} = \frac{1}{2}\frac{v_0 d}{GM}\sigma_v \qquad (1)$$

where $\sigma_{GM}/GM$ is the fractional accuracy of *GM* (where *G* = the Gravitational Constant and *M* = mass of asteroid), *d* is the flyby distance, $v_0$ is the approach speed of the spacecraft at infinity, and $\sigma_v$ is the 1-sigma Doppler noise of the velocity measurement by the RSE. The drift-by distances for a 1% accuracy *GM* determination using the Ka/Ka band with 600 s integration time ($\sigma_v = 5 \times 10^{-4}$ m s$^{-1}$) are given in Table 1.

Once the spacecraft is in a bound orbit, *GM* may be determined more precisely. The driver in the error of *GM* is the uncertainty of the distance to the asteroid. The true mean density follows from the volume determination (from optical observations of the shape) and the mass determination. The mass determination is much more precise than the volume determination; therefore the error of the volume determination is the driver for the error in mean bulk density.

### 3.2 Gravitational Attraction vs. Solar Radiation Pressure Uncertainty

For an asteroid modelled as a triaxial ellipsoid with axes (*a*: *b*: *c*) (with *a* the longest axis) the gravity potential $V(r,\theta,\varphi)$ can be found from the spherical harmonics expansion (Kaula, 1966):

$$V(r,\theta,\phi) = \frac{GM}{r}\left[1 + \sum_{n=2}^{\infty}\sum_{m=0}^{n}\left(\frac{a}{r}\right)^n P_{nm}(\sin\theta)[C_{nm}\cos m\phi + S_{nm}\sin m\phi]\right] \quad (2)$$

where: *r* = distance from Centre of Mass (CoM) of asteroid; $\theta$, $\varphi$ are the co-latitude and longitude of the sub-satellite surface location on the asteroid; *a* = semi-major axis (m); $P_{nm}(\sin\theta)$ = associated Legendre polynomial of order *m* and degree *n*, and $C_{nm}$ and $S_{nm}$ are the gravity coefficients. We can estimate the contribution to the gravitational acceleration at a distance *r* from the CoM for 2002 AT$_4$ and (10302) 1989 ML for different order *m* and degree *n*, for different assumed *GM* and axial ratios (c.f. Table 1). We consider a situation where $\theta$ and $\varphi = 0$. These accelerations are compared in Figure 2 with the estimated Doppler noise (Ka/Ka and X/X) for the RSE for 2002 AT$_4$. We can see that, considering the nominal *GM* case only, it is possible to measure up to order 2 for a 1000 m radius orbit if the axial ratio is (2:1:1), and up to order 4 if it is (3:1:1). For (10302) 1989 ML, *GM* is large enough that for all orbits up to 2000 m we can measure up to degree/order 4.

For the orbital RSE measurements to be successful, all perturbing forces on the spacecraft must be modelled accurately, including the Solar Radiation Pressure (SRP). Non-gravitational perturbations can be difficult to characterise *a priori* because they require detailed modelling of: spacecraft geometry and surface properties; attitude behaviour; the spatial and temporal variations of the incident radiation and particle fluxes, and the interaction of these fluxes with the surfaces. The limitations on the accuracy are likely to be a result of varying SRP over the spacecraft surfaces.

In this study, the DQ Orbiter was based on the SMART-1 chassis, but with larger solar arrays. This was modelled as a $1 \times 1 \times 1$ m$^3$ box with 13.4 m$^2$ total solar panel area, with the optical properties given in Table 4. To estimate the uncertainty in the SRP while in orbit around the target asteroids, we assumed the uncertainty in its attitude to be negligible. We assumed a case where the spacecraft is parallel to the solar direction ($\theta = 0°$), with one side of the spacecraft (i.e. 1 m$^2$, covered in MLI) exposed to the Sun as well as the solar panels. We computed the overall SRP by summing the components from each exposed surface type. In this geometry, the equation given in (Milani et al. 1987) for calculating the radiation pressure acting on a flat plate (assuming Lambertian diffusion) simplifies to:

$$a_{\text{SRP}} = \frac{S_0 A \cos\theta}{r^2 Mc} 2(\delta/3 + \rho) \qquad (3)$$

where $a_{\text{SRP}}$ is the acceleration due to the Solar Radiation Pressure (in the anti-Sun direction): $S_0$ = Solar Constant (1368 W m$^{-2}$), $r$ = Distance from Sun in AU, $A$ = surface area of the flat plate, $\theta$ = angle between surface normal and solar incidence, $M$ = Orbiter mass, $\delta$ = diffusive reflectivity, $\rho$ = specular reflectivity.

The change in $\alpha$ for loaded solar cells between Beginning of Life (BOL) and End of Life (EOL) is determined by the electrical degradation of the solar arrays, which is governed almost entirely by the radiation exposure. The radiation exposure is in turn determined by the radiation environment and the coverslide thickness. As a first-order estimate, the EOL $\alpha$ for loaded solar cells increases by the same amount that the electrical output power decreases and it is common that the difference in BOL and EOL solar array power is 10%. For unloaded cells (cells not providing any current) there is basically no difference in BOL and EOL $\alpha$. All surfaces on SMART-1, with the exception of the solar cell coverslides, can be regarded as diffusely reflective. We unfortunately do not have figures on the specularity of the solar cell coverslides. For the solar panels we can divide the reflection coefficient into specular ($\rho$) and diffusive ($\delta$) coefficients by assuming a specular reflection factor $s$ = 0.30 for the solar cells coverslides [where $\rho = sr$ and $\delta = (1-s)r$]. 80% of the solar array is covered in solar cells. The remaining 20% has the same thermo-optical properties as the rear side of the solar array. For this analysis, we assumed 30% of the solar cells to be loaded. We assume $M$ = 485 kg (with 70 kg of Xe propellant already expended); the uncertainty analysis does not take into account uncertainties in the spacecraft mass.

The optical properties of the spacecraft surfaces will alter as a result of their exposure to the radiation and micrometeorite environment. ESTEC have measured "average" thermo-optical properties of spacecraft materials. It is reported (P. Rathsman, personal communication, 2006) that there is sufficient data to assess the degradation over a mission lifetime to an accuracy of about 10%. We assume that the optical properties of the spacecraft surface materials are known to 1% at BOL and 10% at EOL. We derived $a_{SRP}$ for the minimum and maximum reflection coefficients at BOL and EOL, at perihelion and aphelion for 2002 AT$_4$ and (10302) 1989 ML. The uncertainty in the SRP acceleration due to limitations in our knowledge of the spacecraft surfaces' optical properties, $\Delta a_{SRP}$ (m s$^{-2}$), is roughly the difference between the maximum and minimum accelerations.

We can compare the estimated SRP acceleration noise with the 1$\sigma$ RSE Doppler noise (Figure 3). The uncertainty in acceleration (at EOL) hence comes from assuming that a reflection coefficient ($\delta$) could range from (EOL) $\delta(1-0.1)$ to $\delta(1+0.1)$, while at BOL the range is much smaller, and hence the uncertainty in acceleration is an order of magnitude smaller. What has yet to be investigated is the rate at which ($\Delta\delta$) increases with time, and how to reduce that rate. The uncertainty at BOL is not much greater than the Doppler noise uncertainty, while at the EOL it at least an order of magnitude greater and thus dominates when doing RSE.

Figure 4 shows the SRP acceleration noise compared with the estimated acceleration caused by different order/degree spherical harmonics in the gravity field for 2002 AT$_4$ modelled as a triaxial ellipsoid. For 2002 AT$_4$ (nominal GM, 2:1:1 axis ratio, orbiting at 1000 m radius from CoM) we can only detect up to J2 of the gravity field at EOL, but can detect J4 at BOL. So it is clear that it is important to do the RSE as early as possible. The earlier we arrive at the asteroid after launch the more accurate the RSE measurements. That this is the case makes it important to establish the relationship between [time after launch] and [SRP uncertainty] since this could seriously constrain mission scenarios; here we effectively have two end points and how many years after launch EOL represents is not well constrained.

An important consequence for mission operations is that a campaign to determine the SRP when in orbit may be needed, for example by measuring the change in pressure at different distances from the Sun, or by tilting the solar arrays and measuring the angular acceleration.

## 4     Stereo Mapping Trade-off

In order to achieve the primary objective of measuring semi-major axis change to 1% accuracy, which implies accuracy of order 1 m for the baseline scenario of (10302) 1989 ML, the absolute height resolution must be < 1 m. Even finer relative height resolution would help provide a more accurate shape/mass/density determination and help address secondary objectives that place the asteroid in context. Available instrument types are a camera with stereo coverage, and a laser altimeter.

For both a camera and laser altimeter, the uncertainty in the absolute position of the instrument relative to the asteroid is driven by the SRP uncertainty. Asteroid rendezvous missions have experience of this using both cameras [AMICA on Hayabusa (Hartley & Zisserman 2003), (Demura et al. 2006)] and laser altimetry (NLR on NEAR (Cheng et al. 2002). In both missions the uncertainty in the spacecraft position resulted in ~1 m accuracy of the absolute range measurements.

For relative height resolution, stereo images can achieve much finer *vertical* resolution than laser altimetry. However, most stereo matchers need a correlation patch size. e.g. 10 pixels in diameter – any details within this patch size become blurred. Areas with little texture need a larger correlation patch size, so *spatial* resolution becomes coarser. In smooth terrain, the spatial resolution of stereo imaging may be decreased by several factors. Stereo imaging would have difficulty functioning in this terrain. It is possible that the target NEA could have such terrain covering the majority of its surface. Saito *et al.* (2006) found that 20% of Itokawa's surface is smooth terrain. Additionally, scientifically interesting border regions between smooth and rough terrain may have inaccurate topography. In contrast, laser altimetry does not need texture (or indeed solar illumination) to function. Since the projected mission duration in proximity to the targets is several years, the possible cumulative spatial sampling from laser altimetry, limited only by beam divergence, can give a much finer spatial resolution.

For successful stereo processing, the local solar altitude must not differ by more than 5° and the azimuth by 45°. This places constraints on the imaging campaign and orbit strategy, therefore it would take a lot longer to do global mapping. We estimate that it would take approximately twice as long to acquire a Digital Elevation Model of similar spatial resolution without a laser altimeter. Stereo imaging requires much more data to be downlinked, dedicated ground-based image analysis and complex payload operations based around the requirements for observing the same surface element from different angles under similar illumination conditions. Additionally, laser altimetry has an advantage in that it can work on the night side, and can also obtain topography and shape information from areas in permanent shadow.

A trade-off between using a Mapping Camera alone (assuming the characteristics of AMIE from SMART-1) or having both a Mapping Camera and a Laser Altimeter (AMIE + NLR) is summarised in Table 5. The conclusion of this trade-off is that a laser altimeter should be the primary method of ranging to the surface of the asteroid for orbit determination (primary objective) and determining the global shape model.

## 5 Thermal IR Spectrometer: measuring the Yarkovsky Effect

The Yarkovsky effect is described in detail by Bottke *et al.* (2002) and references therein. We will assume the spin axis of the target asteroid is perpendicular to the orbital plane, i.e. the obliquity $\zeta = 0°$. In this case the asteroid's semi-major axis is only affected by the diurnal Yarkovsky effect. If the asteroid has a prograde rotation, there is a secular increase in the semi-major axis of the asteroid, while if it is retrograde there is a decrease. The magnitude of the diurnal effect depends on Sun-asteroid distance, the subsolar latitude, the size, shape, rotation rate, and surface thermal inertia. In order to achieve the primary objective to acceptable accuracy, the Yarkovsky effect must be measured. This will be done by the RSE in the months before the impact, since any change in semi-major axis beyond those caused by well-understood orbital dynamics must be due to the Yarkovsky effect. However, decoupling uncertainties in the RSE with changes due to the Yarkovsky effect will only be possible with an independent estimation that can come from modelling of the expected effect after mapping of the surface temperature.

We can estimate the diurnal Yarkovsky effect, following the formulation given by Vokrouhlický et al. (2000) for a circular orbit, and assuming the nominal diameter, albedo, rotation period and density given in Table 1 for both objects. Figure 5 shows how the effect varies with thermal inertia. We find the possible Yarkovsky drift could range between 35 and 220 m/yr for 2002 $AT_4$ considering the possible range of thermal inertia assumed in Table 1. For (10302) 1989 ML the range is between 50 and 130 m/yr. Since this is over an order of magnitude larger than the required 10 m semi-major axis change accuracy (over a measurement campaign of comparable length) we conclude that a thermal IR spectrometer is mandatory for achieving the Primary Objective of the DQ mission.

We can define a requirement that the Yarkovsky effect must be measured so that the uncertainty in its semi-major axis drift is less than 4 m (i.e. contributing 40% to the overall "uncertainty budget" for this measurement). Let us assume that we measure the surface thermal inertia to be $41 \pm 1$ J m$^{-2}$ K$^{-1}$ s$^{-1/2}$, i.e. a low thermal inertia on the steep rise of the curve in Figure 5. We find a range of semi-major axis drift between 179 m/yr and 173 m/yr for 2002 AT$_4$. Therefore we require a thermal IR radiometer/spectrometer to be capable of measuring the surface thermal inertia to an accuracy of around 1 J m$^{-2}$ K$^{-1}$ s$^{-1/2}$.

A trade-off vs. a thermal IR radiometer indicated that high spectral resolution is required to measure surface temperature to the required accuracy. This is due to uncertainties introduced by a radiometer not being able to measure the emission spectrum. There would also be additional inaccuracy from beaming and shadowing that could be modelled and removed if we could measure the spectral shape. Additionally, the spectrometer could derive limits on near-surface particle sizes down to a thermal skin depth of 9 cm and a thermal IR spectrometer with a high spectral resolution can be used to determine silicate mineralogy. A sophisticated thermal modelling study is required to account for the above effects and properly size a future thermal IR spectrometer for a mission like DQ.

## 6      Impact Camera

A consequence of adopting the Impact Interpretation Objective would be the inclusion of a second, Impact Camera, optimised to observe the impact (wide field of view, short exposure time, high sensitivity range), while the other is optimised for mapping the asteroid surface (Mapping Camera).

An intriguing possibility is the inclusion of a polarimetry filter wheel. Polarimetry is essentially a method of sensing the texture of a surface. Polarimetry on an asteroid has never been successfully carried out from a spacecraft rendezvous (Hayabusa conducted a test observation with its AMICA camera, but subsequent observations were cancelled due to attitude control problems), but there is a rich history of ground-based polarimetric observations. Interpretation of polarimetry to determine material properties of asteroid surface relies on comparison with laboratory samples. Unfortunately, interpretation of the polarimetric parameters such as depth of negative polarization $P_{min}$, polarisation slope $h$, and inversion angle $α_{inv}$ [for the definition of these parameters see Dollfus *et al.* (1989)] into unambiguous material properties is often unclear. However, polarimetry of the asteroid surface could supply approximate limits on the surface grain size distribution and roughness.

If two cameras are carried, the polarimetry filter could be employed on one camera during the impact to measure the grain size distribution of the impact plume. This is easier to interpret than polarimetry of the asteroid surface, and there is a considerable heritage of interpreting polarimetry of cometary plumes (e.g. Das et al. 2004, and references therein). Understanding the impact grain size distribution will be invaluable for interpreting the results of the impact experiment, enabling the impact to be contextualised for the wide range of possible asteroid surfaces. Therefore the inclusion of a second camera is rated as high priority.

In order to size the Impact Camera, we have produced a simple outflow model which assumes that DQ forms a 10 m diameter hemispherical crater in a surface of density $\rho = 2000$ kg m$^{-3}$ (Figure 6). The ejecta is assumed to evenly fill a constantly expanding 45° cone with particles: we consider the case with all particles of 2 μm, 20 and 200 μm diameter $d_p$. The number of particles in the ejecta, $N$, is calculated from:

$$N = \frac{6M}{\pi d_p^3 \rho} \quad (6)$$

where $M$ is the ejecta mass. We can then calculate the particle density, and consequently the number of photons reflected off the impact cloud per unit area, and finally the S/N of the camera. Figure 7 shows the S/N calculated for different cone heights for 0.05 s exposure times. We can see that if we define that ejecta imaged with S/N>10 can obtain useful science, then a field-of-view (FoV) covering 27.8 km (27.5 + asteroid diameter) is needed. For a 30 km stand-off distance, the Impact Camera should have a FoV of 40°. Measuring the height reached by the ejecta, combined with an estimate of the surface density, can be used to obtain the shear strength around rim of final crater, as was done for the Deep Impact mission. (A'Hearn et al. 2005).

## 7      Payload Operations

We identified several distinct phases of payload operations for the Orbiter after rendezvous. These included:

**Initial RSE drift-bys (7-21 days)**

Several drift-bys (<10) would be performed over a period of between 7 and 21 days. Their purpose is twofold:

(1) to perform RSE during the drift-bys to increase the accuracy of the *GM* estimate to under 10%. It can be seen from Table 1 (×10 for 10% *GM* accuracy drift-by distance) that a 10% accuracy in *GM* can be obtained from a drift-by of 9.4 km for 2002 AT$_4$ and as far out as 87 km for (10302) 1989 ML. Therefore one might envisage an accuracy of near 1% and even J2 terms could be obtained for (10302) 1989 ML.

(2) To improve the global shape model through stereo imaging and laser altimetry. A vertical resolution of ~1 m should be obtainable at 10 km altitude (*d*) from LIDAR ranging.

**Global Mapping (30-60 days, followed by ~300 days of few spectra/images a month)**

An insertion into a metastable orbit of ~1 km altitude is carried out. The orbit should be designed such that the entire asteroid surface can be imaged with the Orbiter Cameras with the solar phase angle α (i.e. Sun-asteroid surface-s/c angle) between 0° and 60°. This orbit should be stable on a timescale of ~1 week, so that a small ΔV for orbit correction can be applied every ~3 days to prevent the spacecraft impacting the asteroid.

The asteroid is globally mapped from 1 km altitude orbit. The theoretical duration of global mapping is determined by the range of possible pole orientations such that the entire surface could only be illuminated by the Sun in anything between 6 hours – 345 days for 2002 AT4 and 19 hours – 210 days for (10302) 1989 ML after the start of this phase. Coverage calculations show that for a pole orientation at 90° to the ecliptic (i.e. with the entire asteroid surface illuminated once per rotation, global mapping with the Orbiter Cameras, NIR spectrometer and Thermal IR Spectrometer is possible in (~60 days). For a pole orientation at 60° to the ecliptic 90% of the surface can be imaged by the Global Mapping Camera in 30 days, but after 90 days, 94% of the surface is mapped (Figure 8). Therefore the strategy is to globally map for ~2 months, and then take a "trickle" of images as more of the surface become available for imaging. The X-ray spectrometer will not globally map the asteroid, but be operational continuously to acquire as long an integrated exposure as possible due to the expected low X-ray flux at the asteroid. The laser altimeter does not need to have completed globally mapping the asteroid to acquire topography and an improved shape model during this phase, since it can continue to cover the surface during RSE phases. The NIR spectrometer would drive the duration of the global mapping phase, since its narrow FoV means it would take 3 months for 2002 AT4 and 6.5 months for (10302) 1989 ML. Low resolution global NIR spectrometer could be taken in drift-bys, with only higher resolution in selected areas of the surface, although there is ample time in the mission timeline for full global mapping.

**RSE campaign (180 days)**

The Orbiter will assume a parking orbit ~50 km from the asteroid for a duration of 230-515 days for 2002 $AT_4$ and 140-290 days for (10302) 1989 ML (depending on the duration of the Detailed Surface Imaging Phase, which in turn depends on the asteroid's pole orientation). All instruments will be non-operational, except occasionally in calibration modes.

The Orbiter will manoeuvre into a 1 km sun-synchronous terminator orbit for RSE operations. to measure the asteroid's semi-major axis and to obtain higher order gravity field harmonics. The projected duration is 180 days. The laser altimeter and the Global Mapping Camera will be operational also. Due to the SRP, the orbit will be at a high solar phase angle, so much of the asteroid surface will be in darkness.

**Impact Monitoring (14 days)**

One week before the impact the Orbiter will move to a ~30 km (TBD) parking orbit to observe the impact. For optimum illumination conditions the position of the Orbiter should be chosen such that the shadow of the impact can be observed on the asteroid surface (approximately 45°-60° Sun-impact site-Orbiter angle). This will allow the velocity of the impact plume to be effectively measured from another angle. The impact will be imaged by both the Impact Camera and the Global Mapping Camera. For portions of the observation, the Impact Camera will use a polarimetry filter. NIR and Thermal IR spectroscopy can be performed on the ejecta cloud.

**Impact Crater Imaging (30 days)**

The Orbiter will be reinserted into a ~1 km altitude surface imaging orbit similar to the initial Detailed Surface Imaging phase after the ejecta has settled onto the asteroid surface or escaped from orbit. The impact crater (~10 m diameter) will be imaged and also the surrounding fresh ejecta, which might cover a significant (i.e. >50%) fraction of the asteroid surface. All the instruments will be operational.

**Second RSE campaign (180 days)**

The Orbiter will be re-inserted into a 1 km altitude sun-synchronous terminator orbit. RSE will be performed to measure the semimajor axis change. Also the laser altimeter will be employed to measure the asteroid's pole orientation in case a change is measurable as a result of the impact (Orbiter Camera imaging would require off-nadir pointing of the camera).

## 8    Conclusions

Many important lessons were learnt during the course of this study. These include:
- Solar radiation pressure uncertainty due to degradation of spacecraft optical surfaces: For low mass asteroids such as 2002 $AT_4$, to measure harmonics above J2, one must do RSE drift-bys for gravity field as early as possible, or perform measurement of SRP in orbit and/or in cruise.

- To fulfil the primary mission objective, a sensitive thermal infrared spectrometer is required to measure the Yarkovsky effect. More complex modelling is required to determine performance requirements.

- A laser altimeter has many advantages and aids achieving the primary objective (increased spatial resolution, smaller data volume than trying to do the same with imagery, fewer operational constraints, etc.).

- To interpret the impact and extrapolate for any particular NEA, via assessment of properties of first few metres (~ crater depth), the payload needs to include:

    – an impact camera (modelling ejecta dynamics required)

    – a polarimetry filter on at least one of the cameras

Future NEO impact mitigation demonstration missions will need to take into account the issues highlighted by this work.

## Acknowledgements

This work was funded under ESA contract No. 4956. We are grateful to the help and advice of Andrés Galvez and Ian Carnelli (ESA). Prime Contractor: QinetiQ. Subcontractors: Swedish Space Corporation, SciSys, Verhaert Space, The Open University. The AMIE / NLR trade-off was developed with advice from Dr. A. Cook (University of Nottingham, Vision and Image Processing Group). We thank an anonymous reviewer for helpful comments that improved the quality of this manuscript.

ACCEPTED MANUSCRIPT18

Bottke, W. F., D. Vokrouhlický, D. P. Rubincam & M. Brož 2002. in. ed. W. F. Bottke, A. Cellino, P. Paolicchi and R. P. Binzel. (Tucson: The University of Arizona Press), 395.

Britt, D. T. & G. J. Consolmagno 2003, Meteoritics & Planetary Science, 38, 1161.

Carnelli, I., A. Galvez & D. Izzo Year, in, NASA Workshop: Near-Earth Object Detection, Characterization, and Threat Mitigation (http://www.esa.int/gsp/ACT/doc/MAD/pub/ACT-RPR-MAD-2006-NASANEOWS-DonQuijote.pdf

Cheng, A. F., et al. 2002, Icarus, 155, 51.

Cole, T. D., M. T. Boies, A. S. El-Dinary, A. Cheng, M. T. Zuber & D. E. Smith 1997, Space Science Reviews, 82, 217.

Das, H. S., A. K. Sen & C. L. Kaul 2004, Astronomy & Astrophysics, 423, 373.

Demura, H., et al. 2006, Science, 312, 1347.

Dollfus, A. M., M. Wolff, J. E. Geake, D. F. Lupishko & L. Dougherty 1989. in. ed. R. P. Binzel, T. Gehrels and M. S. Matthews. (Tucson: University of Arizona Press), 594.

Grande, M. 2001, Earth Moon and Planets, 85-6, 143.

Hartley, R. & A. Zisserman 2003, Multiple View Geometry (Cambridge: Cambridge Univ. Press

Holsapple, K. A. 2004. in. ed. M. J. S. Belton, T. H. Morgan, N. H. Samarasinha and D. K. Yeomans. (Cambridge: Cambridge University Press), 113.

Josset, J. L., et al. 2006. in Advances in Space Research, 14.

Kaula, W. M. 1966. *Theory of Satellite Geodesy*. Blaisdell, Waltham, MA.

Keller, H. U., U. Mall & A. Nathues 2001, Earth Moon and Planets, 85-6, 545.

Milani, A., A. M. Nobili & P. Farinella 1987, Non-Gravitational Perturbations and Satellite Geodesy (Bristol, UK: Adam Hilger

Milani, A., et al. (2003). Near Earth Objects Space Mission Preparation: Don Quijote Mission Executive Summary. 2010.

Pätzold, M., A. Wennmacher, B. Hausler, W. Eidel, T. Morley, N. Thomas & J. D. Anderson 2001, Astronomy & Astrophysics, 370, 1122.

Pätzold, M., B. Häusler, J. Selle & W. Eidel 2005 , Don Quijote - Radio Science Experiment, ESA Technical Note (ACT-TN-060120-4210-DQ RSE).

Rathke, A. & D. Izzo Year, in, Near Earth Objects, our Celestial Neighbors: Opportunity and Risk (IAU Sumposium 236), ed. A. Milani, G. Valsecchi and D. Vokrouhlický (Cambridge University Press)

Saito, J., et al. 2006, Science, 312, 1341.

Vokrouhlický, D., A. Milani & S. R. Chesley 2000, Yarkovsky Effect on Small near-Earth Asteroids: Mathmatical Formulation and Examples, Icarus 148, 118-138.

## Tables

Table 1: Assumed Physical Properties of DQ Target NEAs used in Study, and (in last two rows) 1% accuracy GM determination for Ka/Ka link 600 s integration time

| | 2002 AT$_4$ | | | (10302) 1989 ML | | |
|---|---|---|---|---|---|---|
| **Physical Properties** | **Min.** | **Nominal** | **Max.** | **Min.** | **Nominal** | **Max.** |
| Taxonomic type | | D | | | X | |
| Absolute visual magnitude: $H$ (mag.) | 20.8 | 21.3[b] | 21.8 | 19.15 | 19.35[b] | 19.55 |
| Phase parameter: $G$ | 0 | 0.15[a] | 0.4 | 0 | 0.15[a] | 0.4 |
| Reduced lightcurve amplitude (mag.) | 0 | 0.29[a] | 2 | | 0.84 | |
| Geometric albedo: $p_v$ | 0.05[b] | 0.07[a] | 0.1[b] | 0.05[b] | 0.07[a] | 0.1[b] |
| Diameter: $D$ (m) | 270 | 320 | 380 | 570 | 680 | 800 |
| Bulk density (g/cm$^3$) | 1.3 | 2.0[a] | 2.7 | 1.3 | 2.0[a] | 2.7 |
| Mass (kg) | $6.7\times10^9$ | $3.5\times10^{10}$ | $2.2\times10^{11}$ | $9.4\times10^{10}$ | $3.3\times10^{11}$ | $9.6\times10^{11}$ |
| Thermal Inertia: $\Gamma$ (J m$^{-2}$ K$^{-1}$ s$^{-1/2}$) | 40 | 200[a] | 2200 | 40 | 200[a] | 2200 |
| Axial ratio ($a/b$) | 1 | 2 | 3 | 2.2 | 2.6 | 3 |
| Rotation Period: $P$ (h) | 6 | 6[b] | 6 | 19 | 19 | 19 |
| Rotation State | probably principal axis, but excited (i.e. tumbling) possible | | | principal axis | | |
| Pole Orientation | Unknown | | | Unknown | | |
| Binary? | Unlikely (16% probability) | | | No | | |
| Hill Sphere Radius (km) (at q) | 16.1 | 27.9 | 45.9 | 41.2 | 62.4 | 89.4 |
| Hill Sphere Radius (km) (at Q) | 42.0 | 73.0 | 120 | 54.3 | 82.1 | 117.7 |
| Surface gravit. acceleration (m/s$^2$) | $3.9\times10^{-5}$ | $9.0\times10^{-5}$ | $9.1\times10^{-4}$ | $9.4\times10^{-5}$ | $1.9\times10^{-4}$ | $3.3\times10^{-4}$ |
| Centrifugal acceleration (m/s$^2$) | $2.0\times10^{-7}$ | $1.4\times10^{-5}$ | $1.8\times10^{-4}$ | $2.2\times10^{-6}$ | $2.9\times10^{-6}$ | $3.7\times10^{-6}$ |
| GM (m$^3$s$^{-2}$) | 0.45 | 2.35 | 10.5 | 6.28 | 21.7 | 64.1 |
| $v_0 d$ (m$^2$ s$^{-1}$) | $1.8\times10^5$ | $9.4\times10^5$ | $4.2\times10^6$ | $2.5\times10^6$ | $8.7\times10^6$ | $2.6\times10^7$ |
| Maximum drift-by distance (km) for 1% accuracy GM determination with $v_0 = 1$ m s$^{-1}$ | 0.18 | 0.94 | 4.2 | 2.5 | 8.7 | 26 |

Notes. [a] Assumed, based on typical NEA properties
[b] Assigned by DQ Statement of Work

Table 2. Linkage between mission objectives and payload experiments

| | Objective | Radio Science | Mapping Camera | Thermal IR Spec. | Laser Altimeter | NIR Spec. | Impact Camera | X-Ray Spec. |
|---|---|---|---|---|---|---|---|---|
| Primary | Mass | Y | | | | | | |
| | Semimajor axis change | Y | | Y | | | | |
| | Gravity field | Y | | | Y | | | |
| | Rotation state | | Y | | | | | |
| | Size/Shape | | Y | | Y | | | |
| Impact | Near-surface density | Y | | | | | | |
| | Impact cloud particle size | | | Y | | | Y | |
| | Near-surface porosity limits | | | | | Y | | |
| | Near-surface grain size limits | | Y | Y | | | | |
| | Near-surface shear strength | | | | | | Y | |
| Secon-dary | Topography / Morphology | | Y | | Y | | | |
| | Mineralogy | | | Y | | Y | | |
| | Elemental Composition | | | | | | | Y |

Table 3. Instrument resources for the *Don Quijote* orbiter payload.

| Instrument | Heritage | Mass w/ Margin (kg) | Power w/ Margin (W) | FoV (°) | Key Publication |
|---|---|---|---|---|---|
| Mapping Camera | AMIE (SMART-1) | 2.3 | 2.2 | 5.3 | Josset *et al.* (2006) |
| Thermal IR Spectrometer | MERTIS (BepiColombo) | 2.6 | 9.4 | 4 | Benkhoff *et al.* (2007) |
| Laser Altimeter | NLR (NEAR) | 5.3 | 17.3 | 0.17 | Cole *et al.* (1997) |
| NIR spectrometer | SIR-2 (Chandrayaan-1) | 2.4 | 3.6 | 0.4 | Keller *et al.* (2001) |
| Impact Camera | AMIE (SMART-1) | 2.1 | 2.3 | 40 | Josset *et al.* (2006) |
| X-ray spectrometer | D-CIXS (SMART-1) | 5.0 | 29.4 | 7 | Grande (2001) |
| **TOTAL** | | **19.7** | **60.6** | | |

Table 4: Optical properties of SMART-1 surface coatings used in SRP calculation [P. Rathsman and B. Ljung (SSC), personal communication, 2006].

| Item/Coating | $\alpha$ BOL | $\alpha$ EOL | $\varepsilon$ BOL | $\varepsilon$ EOL |
|---|---|---|---|---|
| Solar panel cells (unloaded) | 0.91 | 0.91 | 0.84 | 0.84 |
| solar panel cells, (loaded) | 0.739 | 0.753 | 0.84 | 0.84 |
| Solar panel rear side | 0.92 | 0.92 | 0.80 | 0.80 |
| Black MLI | 0.86 | 0.93 | 0.86 | 0.86 |

$\alpha$ = absorption coefficient, $\varepsilon$ = emission coefficient. Reflection coefficient $r = 1 - \alpha$.

Table 5: Stereo Mapping Trade-Off, comparing performance using a mapping camera (e.g. AMIE) alone with that using a mapping camera and laser altimeter (e.g. NLR).

| | **AMIE** | **AMIE + NLR** | **Notes** |
|---|---|---|---|
| Relative Height Resolution | 0.1 m | NLR only 0.3 m (NLR + AMIE 0.1 m) | Function of angle change and surface resolution for stereo imaging |
| Absolute Height Resolution | ~1 m | ~ 1 m | i.e. distance to CoM; dominated by SRP and pointing accuracy |
| Spatial Resolution | 0.9 m | 0.2 m | For stereo imaging, ~10 pixels needed as a correlation patch size |
| Smooth Surfaces | Problems for Stereo Matcher: x 3 worse spatial resolution | NLR works fine | e.g. 20% of Itokawa's surface is smooth terrain |
| Overlapping Smooth/ Rough Terrain | Border regions have low precision topography | No problem | For stereo imaging, spatial resolution must be similar within a factor ~2.5 |
| Illumination Constraints | Local solar altitude must not differ by more than 5°; azimuth by 45° | No Constraints | May take approx. twice the time to get DEM with stereo imaging only; more complicated image interpretation; higher manpower/cost |

**Figures**
Fig.1

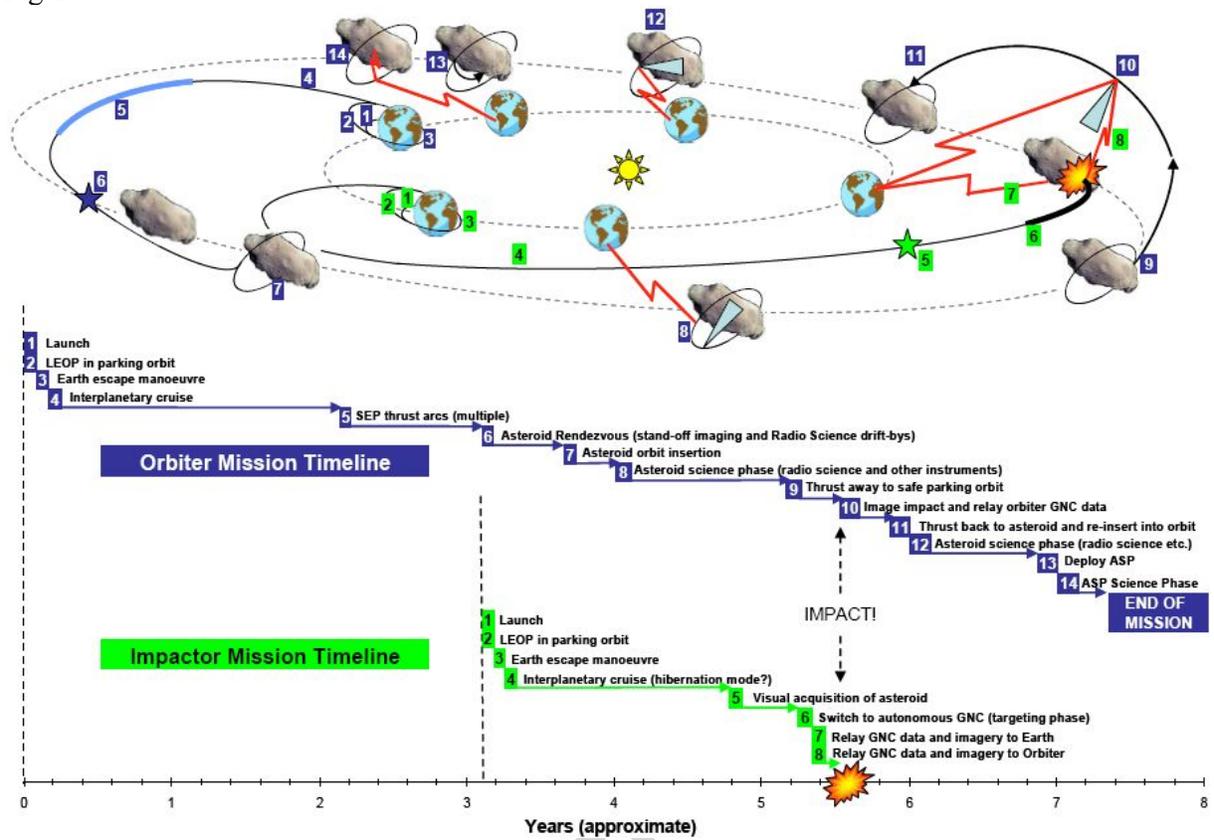

Fig. 2
(a)

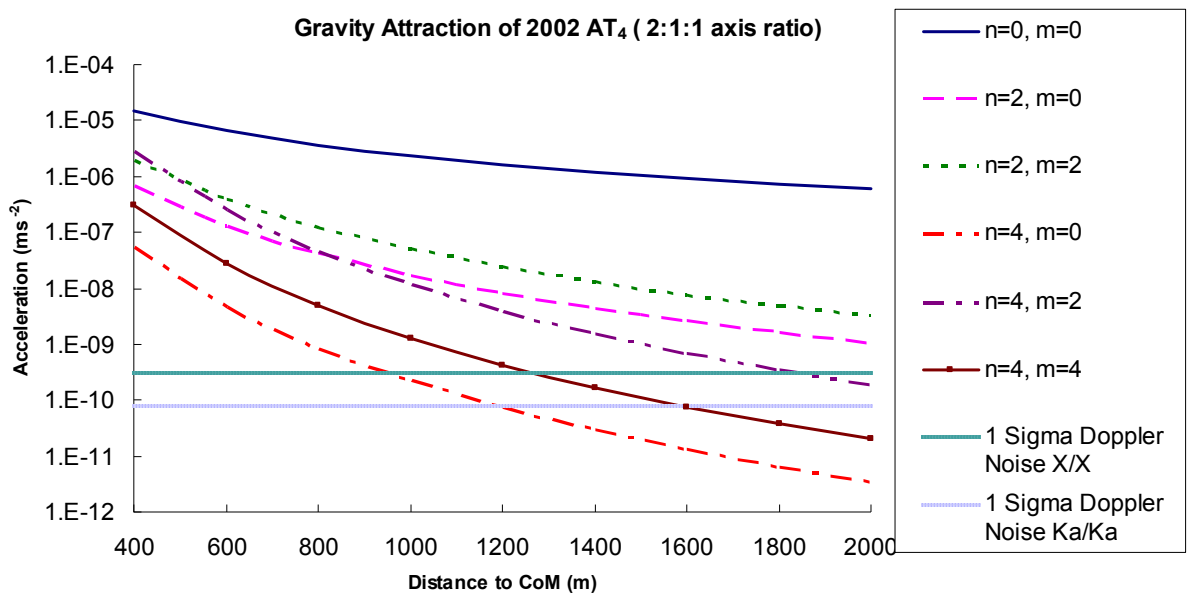

(b)

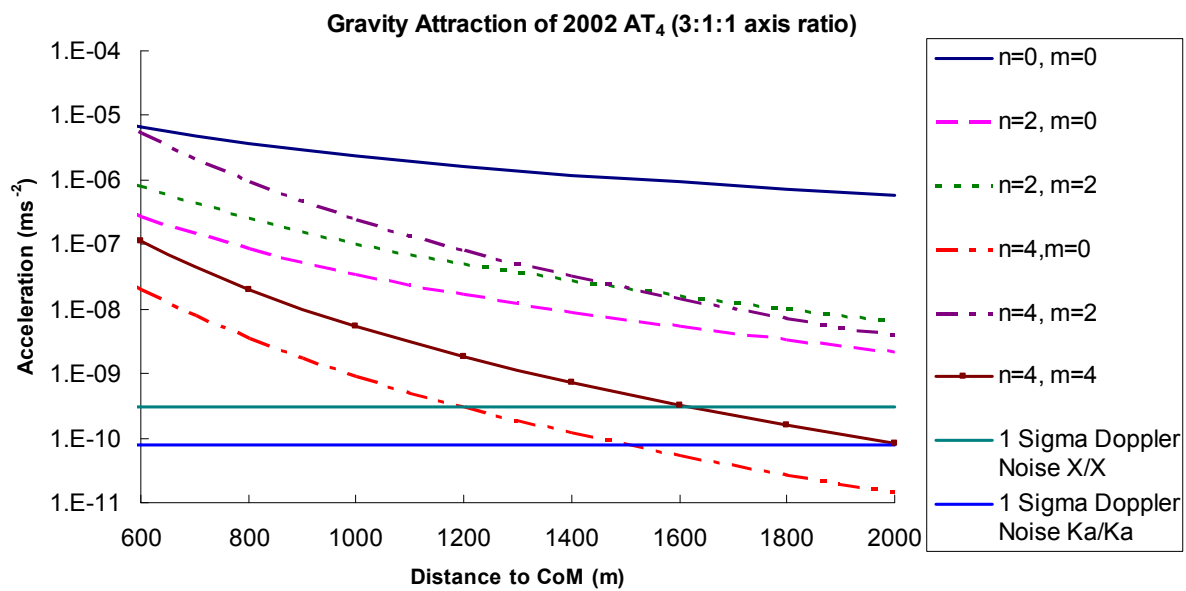

Fig. 3

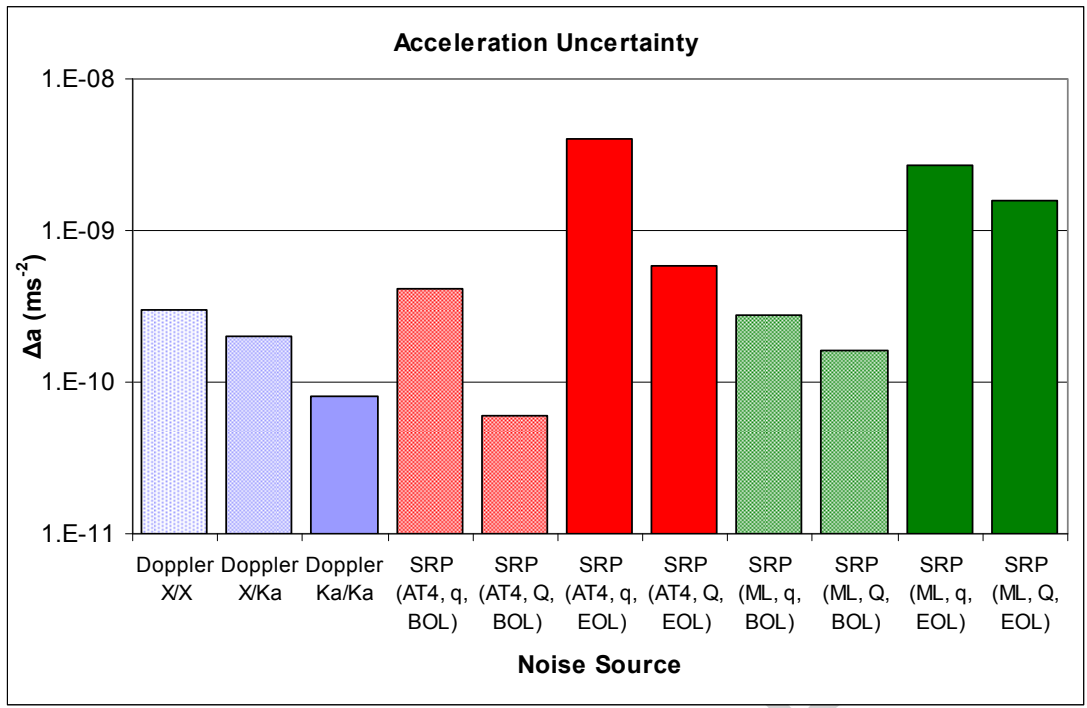

Fig. 4

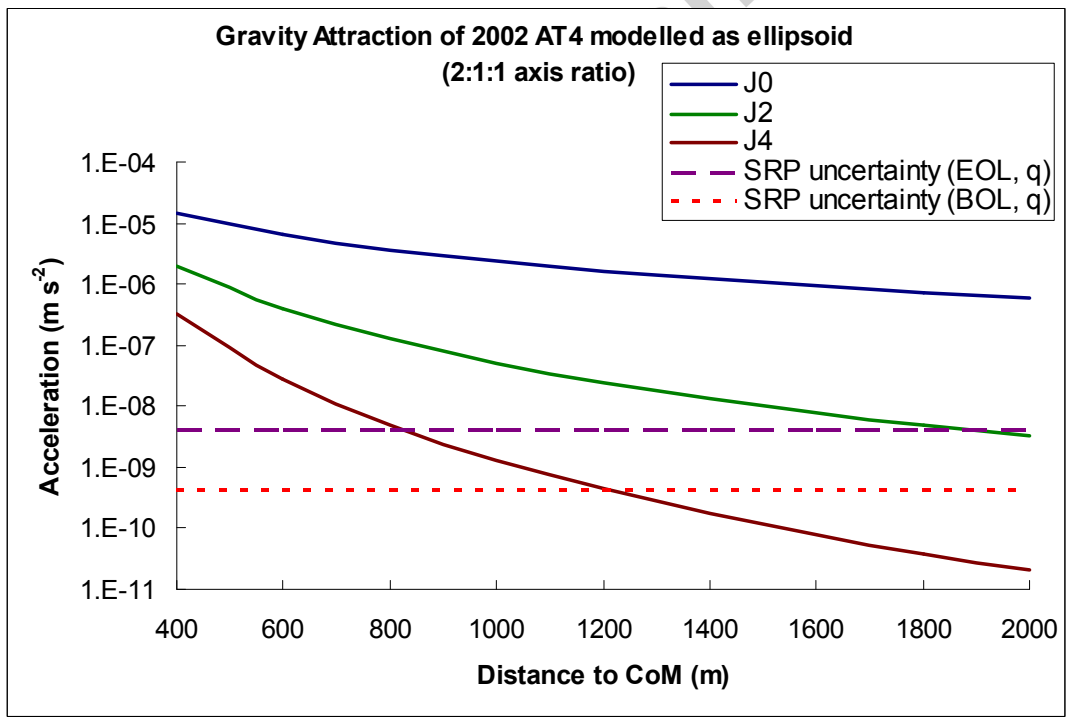

**Fig. 5**

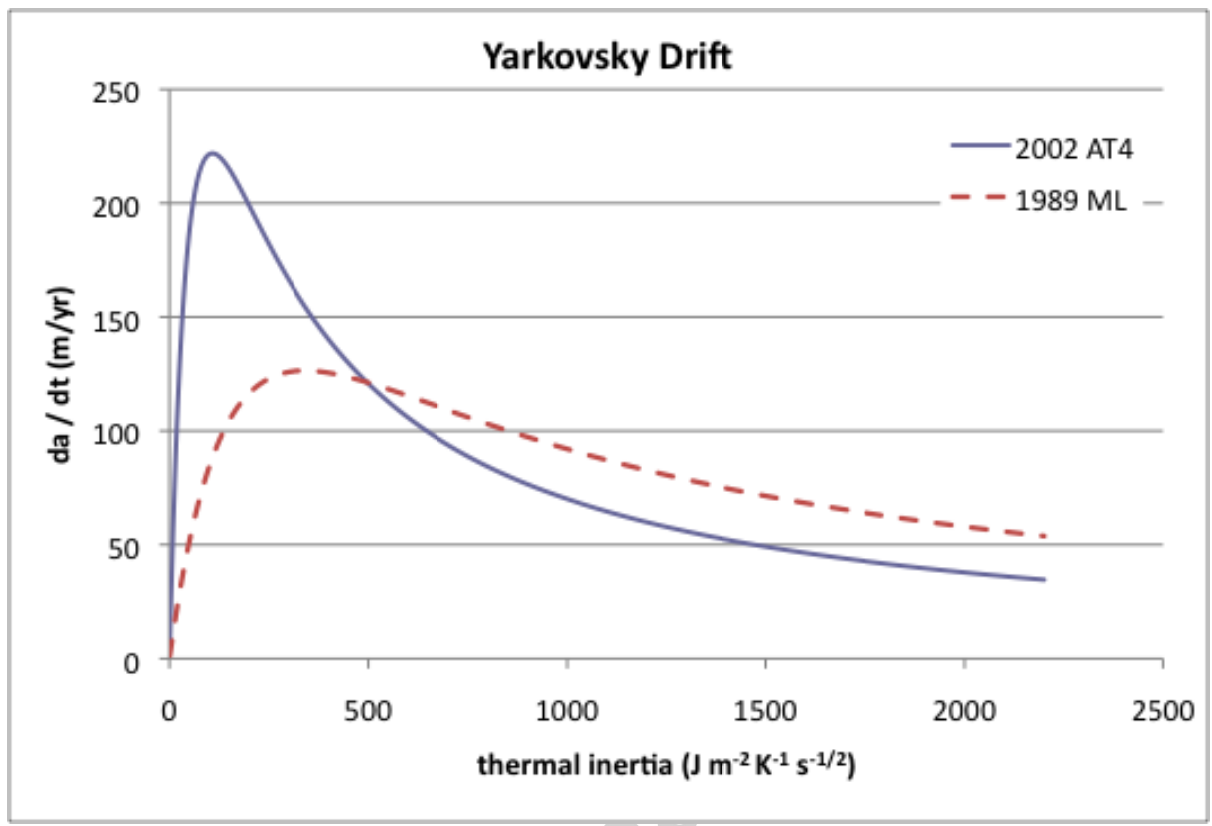

Fig. 6

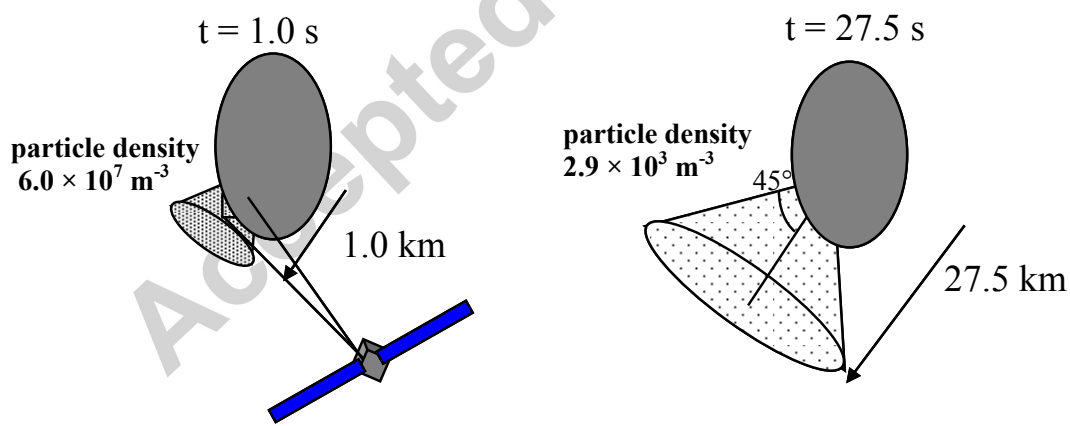

Fig. 7

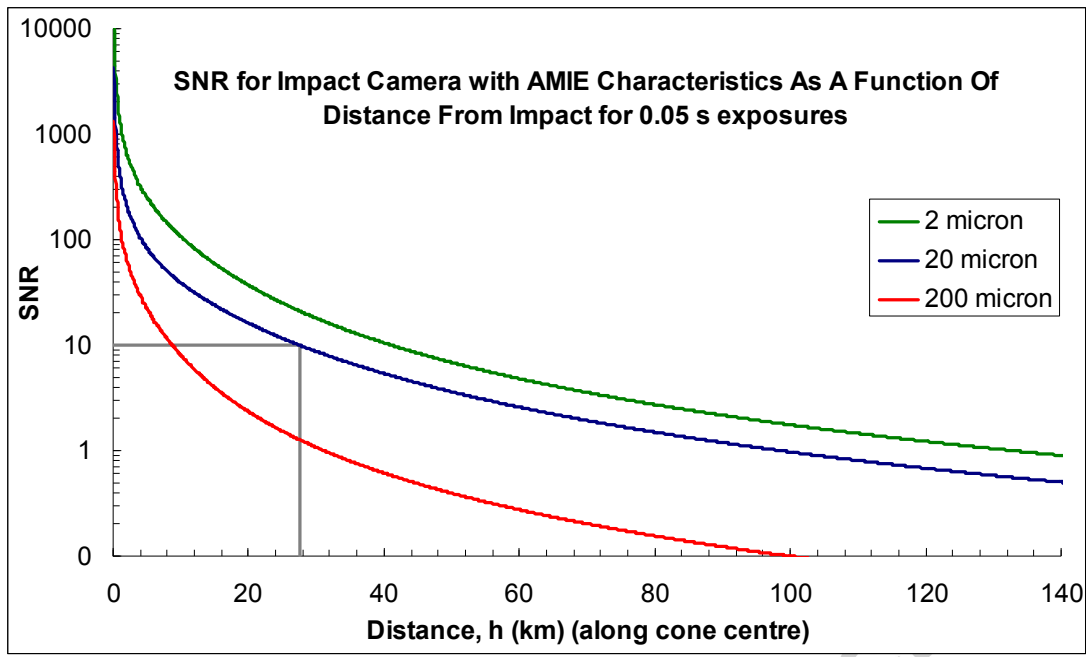

**Fig. 8**

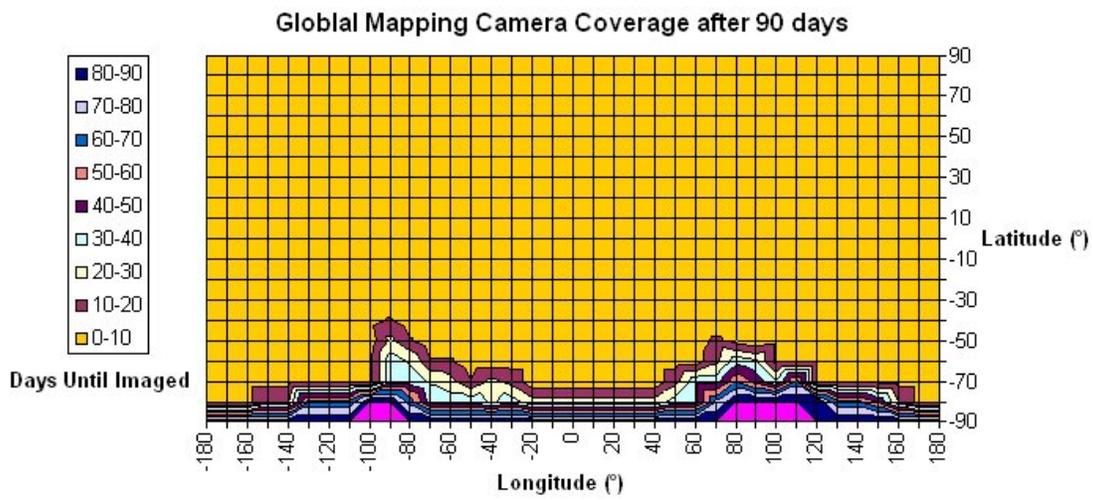

(b)

**Figure Captions**

*Figure 1. Mission profile for the Don Quijote mission, for both Orbiter and Impactor. Reproduced from http://esa-mm.esa.int/docs/NEO/QinetiqDQExecSum.pdf.*

*Figure 2. Contribution to the gravitational acceleration at a distance r from the CoM for 2002 $AT_4$ for different order m and degree n, for assumed nominal GM and different axial ratios compared to $1\sigma$ Doppler noise for 600 s integration time, at latitude 0° and longitude 0° above the asteroid surface. The lines show the 1-sigma Doppler acceleration noise estimated by Pätzold et al. (2005) for X-band uplink (7100 MHz)/X-band downlink (8400 MHz) and Ka-band uplink (34000 MHz)/Ka-band downlink (32000 MHz). The accelerations given are not cumulative, i.e. each curve accounts for the effect of a single coefficient.*

*Figure 3: SRP acceleration noise compared with Doppler noise (600s integration time) for the RSE.*

*Figure 4. SRP uncertainty at Beginning and End of Life (BOL, EOL) for SMART-1 chassis versus gravitational acceleration contribution from different harmonics.*

*Figure 5. Diurnal Yarkovsky drift, change in semi-major axis in metres per year (increase/decrease for prograde/retrograde rotator), estimated for target asteroids as a function of thermal inertia.*

*Figure 6. Simple conical outflow model of DQ impact cloud*

*Figure 7. SNR for camera with AMIE-like characteristics for ejecta cone expanded to different diameters*

*Figure 8: Coverage of 5.3° FoV Global Mapping Camera after 90 days. After 30 days, 90% surface covered); 60 days, 93% surface covered; 90 days, 94% surface covered.*

**Research Highlights for "Measurement Requirements for a Near-Earth Asteroid Impact Mitigation Demonstration Mission"**

Stephen D. Wolters, Andrew J. Ball, Nigel Wells, Christopher Saunders, and Neil McBride

- We present results of a study for a space mission to an asteroid
- Don Quijote: Impactor spacecraft changes asteroid orbit, Orbiter measures change
- Payload is prioritised through trade-offs, taking into account mission objectives
- Orbiter must carry a thermal IR spectrometer, to account for the Yarkovsky effect
- Also discuss Radio Science Experiment, laser altimeter, wide-angle impact camera